\documentclass[10pt]{article}
\usepackage{latexsym,graphicx}
\usepackage{latexsym}
\usepackage{amssymb}
\usepackage{amsmath}
\usepackage{amscd}
\usepackage{amsthm}
\newcommand{\be}{\begin{equation}}
\newcommand{\ee}{\end{equation}}
\usepackage[left=2.5cm,top=2.5cm,right=2.5cm,bottom=1.5cm]{geometry}

\catcode `\@=11
\catcode `\@=12
\begin{document}
\begin{center}
\large{\bf{Anisotropic Bianchi-I Cosmological Model in String Cosmology with Variable Deceleration Parameter}}\\
\vspace{10mm}
\normalsize{Chanchal Chawla$^{1\,\star}$, R. K. Mishra$^{2\, \dagger}$, Anirudh Pradhan$^3$}\\
\vspace{5mm}
\normalsize{$^{1,2}$Department of Mathematics, Sant Longowal Institute of Engineering and Technology,
Longowal$-$148106, Punjab, India \\
\vspace{2mm}
$^{\star}$E-mail: c.chawla137@gmail.com \\
\vspace{2mm}
$^{\dagger}$E-mail: ravkmishra@yahoo.co.in} \\
\vspace{5mm}
\normalsize{$^{3}$Department of Mathematics, Hindu Post-graduate College, Zamania-232 331,
Ghazipur, India \\
E-mail: pradhan@iucaa.ernet.in; pradhan.anirudh@gmail.com} \\
\end{center}
\vspace{10mm}

\begin{abstract}
The present study deals with spatially homogeneous and anisotropic Bianchi type I cosmological model representing massive strings. The energy-momentum tensor, as formulated by Letelier (Phys. Rev. D 28: 2414, 1983) has been used to construct massive string cosmological model for which we assume that the expansion scalar in the model is proportional to one of the components of shear tensor. The Einstein's field equations have been solved by considering time dependent deceleration parameter which renders the scale factor $a = (\sinh(\alpha t))^{\frac{1}{n}}$, where $\alpha$ and $n$ are constants. It has been detected that, for $n > 1$, the presented model universe exhibits phase transition from early decelerated phase to accelerating phase at present epoch while for $0 < n \leq 1$, this describes purely accelerating universe which is consistent with recent astronomical and astrophysical observations. Moreover, some physical and geometric properties of the model along with physical acceptability of the solutions have been also discussed in detail.
\\

\smallskip
\noindent Keywords: String; Bianchi-I metric; Deceleration
parameter; Accelerating universe\\

\smallskip
\noindent PACS Nos. 98.80.K; 04.20.Ex; 04.20.Jb
\end{abstract}

\section{Introduction}
\noindent
The origin of galaxies and formation of other large-scale structures of the Universe are based upon the symmetry
breaking phase transitions after the big bang explosion, when the temperature falls below some critical temperature
as predicted by Grand Unified Theories (GUT) (Zel'dovich et al. 1975; Kibble 1976, 1980; Everett 1981; Vilenkin
1981, 1985). Such type of transitions lead to the formation of {\it topologically stable defects} such as vacuum
domain walls, cosmic strings, monopoles, textures and other `hybrid' creatures. To study the effects of cosmic
strings is one of the most outstanding problems in cosmology as the vacuum strings (Zel'dovich 1980) generate density perturbations which are strong enough for the formation of galaxies.  Moreover, the large scale network of strings
during the early universe does not contradict the present-day observations (Kibble, 1976). These strings also
produce gravitational effects that arise from the coupling between their stress-energy and gravitational field.
The general relativistic treatment of strings was pioneered by (Letetier 1979, 1983) and (Stachel 1980).
(Letelier 1979) formulated the energy-momentum tensor for classical massive strings while (Letelier 1983) represented some cosmological solutions of massive strings in Bianchi type I and Kantowski-Sachs space-times. The cloud of strings are termed as massive strings and each massive string is formed by a geometrical string with particles attached along its extension. Therefore, the cloud forming strings are the generalization of Takabayasi's relativistic model of strings (known as p-strings) where particles and strings are together. Since strings are not observable today, so we can say that the
strings shrink and disappear with the evolution of universe and these cosmic strings end up with a cloud of particles.
Recently, (Pradhan et al. 2007) and (Yadav et al. 2009) prevailed inhomogeneous string cosmological models formed by geometrical strings and used these models as a source of gravitational fields. In recent years, cosmic strings have been studied by several authors (Banerjee et al. 1990; Krori et al. 1990; Roy and Banerjee 1995; Wang 2005, 2006; Bali and Anjali 2006; Bali et al. 2007; Bali and Pradhan 2007; Yadav et al. 2007a, 2007b; Reddy et al. 2007; Pradhan et al. 2008, 2009, 2010, 2011b; Pradhan and Mathur 2008; Rao et al. 2009, 2011; Pradhan 2007, 2009; Belinchon 2009a, 2009b; Rao and Vinutha 2010; Pradhan and Chouhan 2011; Yadav 2012) in Bianchi type space-times in different physical context. \\

The simplest of anisotropic models, which, nevertheless, rather completely describe the anisotropic effects, are Bianchi type I (BI) spatially homogeneous models whose spatial sections are flat but the expansion or contraction rate is directional dependent. The advantages of these anisotropic models are that they have a significant role in the description of evolution of early phase of the Universe and they help in finding more general cosmological models than the isotropic FRW models. Observations by the Differential Microwave Radiometers on NASA's Cosmic Background Explorer recorded anisotropy in various angle scales. It is speculated that these anisotropies hide in their hearts the entire history of the cosmic evolution down to recombination, and they are conceived to be indicative of the Universe geometry and the matter composing the Universe. The theoretical argument (Misner 1968) and the modern experimental data support the existence of an anisotropic phase, which turns into an isotropic one. The isotropy of the present-day universe makes the BI model a prime candidate for studying the possible effects of an anisotropy in the early universe on modern-day data observations.\\

In literature, it is common to use a constant deceleration parameter (Akarsu and Kilinc 2010a, 2010b; Amirhashchi et al. 2011a; Pradhan et al. 2011a) as it suggests a power law for metric function or corresponding quantity. But recent observations from Type Ia Supernovae (SNe Ia) (Riess et al. 1998, 2004; Perlmutter et al. 1999; Tonry et al. 2003; Clocchiatti et al. 2006) and CMB anisotropies (de Bernardis et al. 2000; Hanany et al. 2000; Bennett et al. 2003) revealed that the current universe is not only expanding but also accelerating. Several researchers (Pradhan et al. 2006; Pradhan and Otarod 2006; Pradhan et al. 2012, Yadav 2012) have discussed evolution of the Universe with variable mean deceleration parameter (DP). Moreover, (Pradhan et al. 2012) proposed the law of variation of scale factor as increasing function of time in Bianchi type $VI_{0}$ space-time, which generates a time dependent DP. This law provides explicit form of scale factors governing the Bianchi type $VI_{0}$ universe and facilitates to describe accelerating phase of the Universe.\\

Motivated from the studies outlined above, in this paper, the Einstein's field equations have been solved for massive strings with time dependent DP in Bianchi type I space-time. This scenario facilitates to describe the transition of universe from early decelerated phase to the present accelerating phase. The paper has the following structure. The metric and field equations are presented in Section $2$. In Section $3$, exact solutions to the field equations with cloud of strings have been presented with various physical and geometric properties of the model. Section $4$ proves physical acceptability of the derived solutions and final conclusions are summarized in the last Section $5$.

\section{The Metric and Field Equations}
We consider totally anisotropic Bianchi type I line element, given by
\begin{equation}
\label{eq1} ds^{2} = -dt^{2} + A^{2}dx^{2} + B^{2}dy^{2} + C^{2}dz^{2},
\end{equation}
where the metric potentials $A$, $B$ and $C$ are functions of $t$ only.
This ensures that the model is spatially homogeneous. \\

The Einstein's field equations (in gravitational units $8\pi G = c = 1$)
read as
\begin{equation}
\label{eq2} R_{i}^{j} - \frac{1}{2}Rg_{i}^{j} = T_{i}^{j},
\end{equation}
where $T_{i}^{j}$ is the energy-momentum tensor for a cloud of massive strings
and perfect fluid distribution, given by
\begin{equation}
\label{eq3} T_{i}^{j} = (\rho + p)u_{i}u^{j} + pg_{i}^{j} - \lambda x_{i}x^{j}.
\end{equation}
Here $\rho$ and $p$ are, respectively, the energy density and the isotropic pressure
for a cloud of strings with particles attached to them; $\lambda$ is the string tension
density; $u^{i} = (0,0,0,1)$ is the four velocity of the particles and $x^{i}$ is a
unit space-like vector representing the direction of string. The vectors $u^{i}$ and
$x^{i}$ satisfy the conditions $u_{i}u^{i} = -x_{i}x^{i} = -1, u^{i}x_{i}=0$. Choosing
$x^{i}$ parallel to $\partial/\partial x$, we have $x^{i} = (A^{-1}, 0, 0, 0)$.\\

If particle density of the configuration is denoted by $\rho_{p}$, then
\begin{equation}
\label{eq4} \rho = \rho_{p} + \lambda.
\end{equation}
The Einstein's field equations (\ref{eq2}) for the line element (\ref{eq1}) and energy
distribution (\ref{eq3}), lead to the following set of independent differential equations:
\begin{equation}
\label{eq5} \frac{\ddot{B}}{B} + \frac{\ddot{C}}{C} + \frac{\dot{B}\dot{C}}{BC}
= - p + \lambda,
\end{equation}
\begin{equation}
\label{eq6} \frac{\ddot{C}}{C} + \frac{\ddot{A}}{A} + \frac{\dot{C}\dot{A}}{CA}
= - p,
\end{equation}
\begin{equation}
\label{eq7} \frac{\ddot{A}}{A} + \frac{\ddot{B}}{B} + \frac{\dot{A}\dot{B}}{AB}
= - p,
\end{equation}
\begin{equation}
\label{eq8} \frac{\dot{A}\dot{B}}{AB} + \frac{\dot{B}\dot{C}}{BC} + \frac{\dot{C}\dot{A}}{CA} = \rho.
\end{equation}
The energy conservation equation $T^{ij}_{;j} = 0$, leads to the following expression:
\begin{equation}
\label{eq9} \dot{\rho} + 3(\rho + p)H - \lambda\frac{\dot{A}}{A} = 0,
\end{equation}
with $H$ being the mean Hubble parameter, which for Bianchi type I space-time can be defined as
\begin{equation}
\label{eq10} H = \frac{\dot{a}}{a} = \frac{1}{3}\left(\frac{\dot{A}}{A} +
\frac{\dot{B}}{B} + \frac{\dot{C}}{C}\right) = \frac{1}{3}(H_{1}+H_{2}+H_{3}),
\end{equation}
where $H_{1} = \frac{\dot{A}}{A}$, $H_{2}=\frac{\dot{B}}{B}$ and
$H_{3} = \frac{\dot{C}}{C}$ are directional Hubble factors in the directions of $x$-, $y$-
and $z$-axes respectively. Here an over dot denotes derivative with respect to the cosmic
time $t$. Also, $a$ is the average scale factor of Bianchi type I model written as
\begin{equation}
\label{eq11} a = (ABC)^{\frac{1}{3}}.
\end{equation}
The usual definitions of the dynamical scalars such as the expansion scalar ($\theta$) and the shear
scalar ($\sigma$) are considered to be
\begin{equation}
\label{eq12} \theta = u^{i}_{;i} = \frac{3\dot{a}}{a}
\end{equation}
and
\begin{equation}
\label{eq13} \sigma^{2} = \frac{1}{2}\sigma_{ij}\sigma^{ij} =
\frac{1}{2}\Biggl[\left(\frac{\dot{A}}{A}\right)^{2}+
\left(\frac{\dot{B}}{B}\right)^{2}+\left(\frac{\dot{C}}{C}\right)^{2}\Biggr]-\frac{1}{6}\theta^{2},
\end{equation}
where
$$ \sigma_{ij} = u_{i;j}+\frac{1}{2}(u_{i;k}u^{k}u_{j}+ u_{j;k}u^{k}u_{i})+
 \frac{1}{3}\theta(g_{ij} + u_{i}u_{j}).$$

The anisotropy parameter ($A_{m}$) is defined as
\begin{equation}
 \label{eq14} A_{m} = \frac{1}{3}\sum_{i=1}^{3}\left(\frac{H_{i}-H}{H}\right)^{2}.
\end{equation}
The field equations (\ref{eq5}) $-$ (\ref{eq8}) involve six unknown variables viz. $A$, $B$,
$C$, $p$, $\rho$ and $\lambda$. In order to solve field equations explicitly, we require two
additional constraints relating these variables, which we shall consider in the following section.\\
\section{Solutions of the Field Equations}

Firstly, we define the mean deceleration parameter ($q$) as
\begin{equation}
\label{eq15} q = -\frac{a\ddot{a}}{\dot{a}^{2}} = - \left(\frac{\dot{H} + H^{2}}{H^{2}}\right)
 = b(t) ~ ~\mbox{say}.
\end{equation}
The motivation to choose such time dependent DP is behind the fact that the Universe exhibits phase transition from the past decelerating expansion to the recent accelerating one, as revealed by the recent observations of SNe Ia (Riess et al. 1998, 2004; Perlmutter et al. 1999; Tonry et al. 2003; Clocchiatti et al. 2006) and CMB anisotropies (de Bernardis et al. 2000; Hanany et al. 2000; Bennett et al. 2003). In their preliminary analysis, it was found that the SNe data favor recent acceleration ($z < 0.5$) and past deceleration ($z > 0.5$). From the cosmological observations in the literature (Cunha and Lima 2008; Cunha 2009; Li et al. 2011; Frieman et al. 2008; Melchiorri et al. 2007; Ishida et al. 2008; Pandolfi 2009; Lima et al. 2010), the transition redshift ($z_{t}$) from decelerating to accelerating expansion is given by $0.3 < z_{t} < 0.8$. More recently, the High-Z Supernova Search (HZSNS) team have obtained $z_{t} = 0.46 \pm 0.13$ at $(1\;\sigma)$ c.l. (Riess et al. 2004) which has been further improved to  $z_{t} = 0.43 \pm 0.07$ at $(1\;\sigma)$ c.l. (Riess at al. 2007). The Supernova Legacy Survey (SNLS) (Astier et al. 2006), as well as the one recently compiled by (Davis et al. 2007), yield $z_{t} \sim 0.6$ $(1\; \sigma)$ in better agreement with the flat $\Lambda$CDM model ($z_{t} = (2\Omega_{\Lambda}/\Omega_{m})^{\frac{1}{3}} - 1 \sim 0.66$). Now for a universe which was decelerating in past and accelerating at the present time, the DP must show signature flipping (Riess et al. 2001; Padmanabhan and Roychowdhury 2003; Amendola 2003). At present, one may examine the variation of cosmic acceleration, i.e., deceleration parameter (DP) with time instead of knowing only the beginning of cosmic acceleration and the present value of $q$. Thus, our choice of variable $q$ is physically acceptable.\\

Equation (\ref{eq15}) may be rewritten as
\begin{equation}
\label{eq16} \frac{\ddot{a}}{a} + b\frac{\dot{a}^{2}}{a^{2}} = 0.
\end{equation}
In order to solve the Eq. (\ref{eq16}), we assume $b = b(a)$. It is important to note here that one can assume
$b = b(t) = b(a(t))$, as $a$ is also a time dependent function.\\

The general solution of Eq. (\ref{eq16}) with the assumption $b = b(a)$, is obtained as
\begin{equation}
\label{eq17} \int e^{\int\frac{b}{a}da}da = t+k,
\end{equation}
where $k$ is an integrating constant. One cannot solve (\ref{eq17}) in general as $b$ is variable. So, in order to solve the problem completely, we have to choose $\int\frac{b}{a}da$ in such a manner that (\ref{eq17}) be integrable without any loss of generality. Hence we consider
\begin{equation}
\label{eq18} \int\frac{b}{a}da = \ln f(a),
\end{equation}
which does not affect the nature of generality of solution. Hence from (\ref{eq17}) and (\ref{eq18}),
we obtain
\begin{equation}
\label{eq19} \int f(a)da = t + k.
\end{equation}
Of course the choice of $f(a)$ in (\ref{eq19}) is quite arbitrary, but for the sake of
physically viable models of the Universe consistent with observations, we here consider
\begin{equation}
\label{eq20} f(a) = \frac{na^{n-1}}{\alpha \sqrt{1+a^{2n}}},
\end{equation}
where $\alpha$ is an arbitrary constant and $n$ is a positive constant. In this case, on integrating Eq.
(\ref{eq19}) and neglecting the integration constant $k$, we obtain the exact solution as
\begin{equation}
\label{eq21} a(t) = (\sinh(\alpha t))^{\frac{1}{n}}.
\end{equation}
This relation (\ref{eq21}) generalizes the value of scale factor obtained by (Pradhan et al. 2012)
in connection with the study of dark energy models in Bianchi type $VI_{0}$ space-time and the one suggested by
(Amirhashchi et al. 2011b) in order to study the evolution of dark energy models in FRW universe filled with a mixture
of barotropic fluid and dark energy by considering time dependent $q$. \\

Secondly, we assume that the component $\sigma_{~1}^{1}$ of the shear tensor ($\sigma_{~i}^{j}$) is proportional
to the expansion scalar ($\theta$), i.e., $\sigma_{~1}^{1} \propto \theta$. This condition leads to
\begin{equation}
\label{eq22} \frac{1}{3}\left(\frac{{3\dot{A}}}{A}-\frac{\dot{B}}{B}-\frac{\dot{C}}{C}\right)
= \beta \left(\frac{\dot{A}}{A}+\frac{\dot{B}}{B}+\frac{\dot{C}}{C}\right),
\end{equation}
which yields
\begin{equation}
\label{eq23} \frac{\dot{A}}{A} = m\left(\frac{\dot{B}}{B}+\frac{\dot{C}}{C}\right),
\end{equation}
where $m = \frac{1+3\beta}{3(1-\beta)}$ with $\beta$ being the constant of proportionality. Above
equation, after integration, reduces to
\begin{equation}
\label{eq24} A = k_{1}(BC)^{m},
\end{equation}
where $k_{1}$ is an integrating constant. Here, for simplicity and without any loss of generality,
we assume $k_{1} = 1$. Hence we have
\begin{equation}
\label{eq25} A = (BC)^{m}.
\end{equation}
The motivation behind assuming this condition is explained with reference to (Thorne 1967), the observations of
the velocity-red-shift relation for extragalactic sources suggest that the Hubble expansion of the Universe
is isotropic today within $\thickapprox 30$ percent (Kantowski and Sachs 1966; Kristian and Sachs 1966). To put
more precisely, redshift studies place the limit $\frac{\sigma}{H} \leq 0.3$ in the neighborhood of our galaxy today. (Collins et al. 1980) have also pointed out that for spatially homogeneous metric, the normal congruence to the homogeneous expansion satisfies the condition that $\frac{\sigma}{\theta}$ is constant. \\

Subtracting (\ref{eq6}) from (\ref{eq7}), and taking integral of the resulting equation
two times, we get
 \begin{equation}
\label{eq26} \frac{B}{C} = c_{1}\exp \left[c_{2}\int (ABC)^{-1} dt\right],
\end{equation}
where $c_{1}$ and $c_{2}$ are constants of integration.\\

Solving (\ref{eq11}), (\ref{eq25}) and (\ref{eq26}), we obtain the metric functions as
\begin{equation}
\label{eq27} A(t) = a^{\frac{3m}{m+1}},
\end{equation}
\begin{equation}
\label{eq28} B(t) = \sqrt{c_{1}} a^{\frac{3}{2(m+1)}} \exp\left[\frac{c_{2}}{2}
\int\frac{1}{a^{3}}dt\right],
\end{equation}
\begin{equation}
\label{eq29} C(t) = \frac{1}{\sqrt{c_{1}}} a^{\frac{3}{2(m+1)}} \exp\left[-\frac{c_{2}}{2}
\int\frac{1}{a^{3}}dt\right].
\end{equation}
Using the relation (\ref{eq21}) in Eqs. (\ref{eq27})$-$(\ref{eq29}), we obtain following expressions for the scale factors:
\begin{equation}
\label{eq30} A(t) = (\sinh (\alpha t))^{\frac{3m}{n(m + 1)}},
\end{equation}
\begin{equation}
\label{eq31} B(t) = \sqrt{c_{1}} (\sinh (\alpha t))^{\frac{3}{2n(m+1)}}\exp{\left[\frac{c_{2}(-1)^{\frac{n+3}{2n}}}{2\alpha}
\cosh(\alpha t) F(t)\right]},
\end{equation}
\begin{equation}
\label{eq32} C(t) = \frac{1}{\sqrt{c_{1}}}(\sinh (\alpha t)) ^{\frac{3}{2n(m+1)}}\exp{\left[\frac{c_{2}(-1)^{\frac{3(n+1)}{2n}}}{2\alpha}
\cosh(\alpha t) F(t)\right]},
\end{equation}
where
\begin{equation}
\label{eq33}
F(t) = 1 + \frac{1}{6}\left(1+\frac{3}{n}\right)\cosh^{2}(\alpha t) + \frac{3}{40}\left(1+\frac{3}{n}\right)
\left(1+\frac{1}{n}\right)\cosh^{4}(\alpha t) + O[\cosh (\alpha t)]^{6}.
\end{equation}
From Eqs. (\ref{eq30})$-$(\ref{eq32}), it is clearly observable that the spatial scale factors are zero at the initial epoch
$t = 0$, indicating that the model has a point type singularity (MacCallum 1971) with the following form:
\[
ds^{2} =  -dt^{2} + (\sinh (\alpha t))^{\frac{6m}{n(m + 1)}}dx^{2} + c_{1}(\sinh (\alpha t))^{\frac{3}{n(m+1)}}\exp{\left[\frac{c_{2}(-1)^{\frac{n+3}{2n}}}{\alpha}
\cosh(\alpha t) F(t)\right]}dy^{2} +
\]
\begin{equation}
\label{eq34}
\frac{1}{c_{1}}(\sinh (\alpha t)) ^{\frac{3}{n(m+1)}}\exp{\left[\frac{c_{2}(-1)^{\frac{3(n+1)}{2n}}}{\alpha}
\cosh(\alpha t) F(t)\right]}dz^{2}.
\end{equation}

For the derived model, the expressions for observational physical quantities such as spatial volume scale factor ($V$), directional Hubble parameters ($H_{i}$), Hubble parameter ($H$), expansion scalar ($\theta$), shear scalar ($\sigma$) and mean anisotropy parameter ($A_{m}$) are given by
\begin{equation}
\label{eq35} V = ABC = (\sinh(\alpha t))^{\frac{3}{n}},
\end{equation}
\[
H_{1} = \frac{3m\alpha}{n(m + 1)}\coth(\alpha t),
\]
\[
H_{2} = \frac{3\alpha}{2n(m + 1)}\coth(\alpha t) + \frac{c_{2}}{2(\sinh(\alpha t))^{\frac{3}{n}}},
\]
\begin{equation}
\label{eq36} H_{3} = \frac{3\alpha}{2n(m + 1)}\coth(\alpha t) - \frac{c_{2}}{2(\sinh(\alpha t))^{\frac{3}{n}}},
\end{equation}
\begin{equation}
\label{eq37} \theta = 3H = \frac{3\alpha}{n}\coth(\alpha t),
\end{equation}
\begin{equation}
\label{eq38} \sigma^{2} = 3\left(\frac{\alpha (2m - 1)}{2n(m + 1)}\coth(\alpha t)\right)^{2} +
\left(\frac{c_{2}}{2(\sinh(\alpha t))^{\frac{3}{n}}}\right)^{2},
\end{equation}
\begin{equation}
\label{eq39} A_{m} = \frac{1}{2}\left(\frac{2m - 1}{m + 1}\right)^{2} +
\frac{1}{6}\left(\frac{nc_{2}}{\alpha \coth(\alpha t)(\sinh(\alpha t))^{\frac{3}{n}}}\right)^{2}.
\end{equation}
From the above Eqs. (\ref{eq35})$-$(\ref{eq39}), it is observed that at $t = 0$, the spatial volume scale factor ($V$)
vanishes while the parameters such as scalar of expansion ($\theta$), mean Hubble parameter ($H$) and shear scalar ($\sigma$) are infinite, which is big bang scenario. As $t \to \infty$, $V$ diverges to $\infty$ whereas $\theta$, $H$ and $\sigma$ approach to zero. Moreover, from Eqs. (\ref{eq37}) and (\ref{eq38}) we obtain
\begin{equation}
\label{eq40}
\frac{\sigma^{2}}{\theta^{2}} = \frac{1}{3}\left(\frac
{2m-1}{2(m+1)}\right)^{2}+\left(\frac{nc_{2}\tanh(\alpha t)}{6\alpha(\sinh(\alpha t))^{\frac{3}{n}}}\right)^{2}.
\end{equation}
Clearly $\frac{\sigma^{2}}{\theta^{2}} \rightarrow \frac{1}{3}\left(\frac{2m-1}{2(m+1)}\right)^{2}$ as $t \rightarrow\infty$
indicating that $\frac{\sigma^{2}}{\theta^{2}} \nrightarrow 0$ as $t \rightarrow\infty$ as long as $m \neq \frac{1}{2}$.
Therefore, the dynamics of mean anisotropy parameter ($A_{m}$) depends upon the value of $m$ and for $m = \frac{1}{2}$,
$A_{m} \rightarrow 0$ as $t \rightarrow \infty$ i.e. the model approaches to isotropy at late times.\\

\begin{figure}[ht]
\centering
\includegraphics[width=12cm,height=6cm,angle=0]{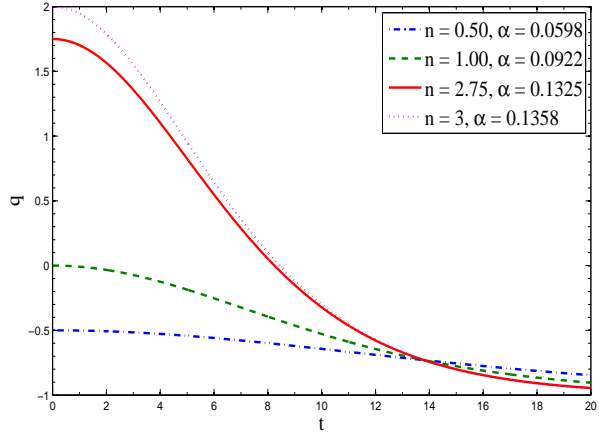}
\caption{The plot of deceleration parameter ($q$) vs time ($t$).}
\label{fig:figure1}
\end{figure}

The mean deceleration parameter ($q$) for the model is found to be
\begin{equation}
\label{eq41} q = n\left(1 - \tanh^{2}(\alpha t)\right) - 1.
\end{equation}
For the present age of the Universe i.e. at $t = t_{0}$, $q = q_{0}$, we get the following relationship between the constants $n$ and $\alpha$:
\begin{equation}
\label{eq42} \alpha = \frac{1}{t_{0}}\tanh^{-1}\left[1-\frac{q_{0}+1}{n}\right]
^{\frac{1}{2}}.
\end{equation}
The sign of $q$ characterizes inflation of the Universe. A positive sign of $q$ i.e. $q > 0$, correspond
to decelerating model whereas negative sign of $q$ (particularly $-1 \leq q < 0$) indicates accelerating phase or inflationary model. From Eq. (\ref{eq41}), we observe that $q > 0$ for $t < \frac{1}{\alpha}\tanh^{-1}(1 - \frac{1}{n})^{\frac{1}{2}}$ and $q < 0$ for $t > \frac{1}{\alpha}\tanh^{-1}(1 - \frac{1}{n})^{\frac{1}{2}}$. Fig. $1$ depicts the behavior of $q$ with time ($t$) for different value of $n$ and $\alpha$ in accordance with the relation (\ref{eq42}) for $t_{0} = 13.8$ GYr and $q_{0} = -0.73$ (Cunha and Lima 2008). It is clearly observable from the figure that for $0 < n \leq 1$, the presented model universe is in accelerating phase whereas for $n > 1$, the model is evolving from decelerating phase to accelerating phase. Also, recent observations of SNe Ia (Riess et al. 1998, 2004; Perlmutter et al. 1999; Tonry et al. 2003; Clocchiatti et al. 2006) and CMB anisotropies (de Bernardis et al. 2000; Hanany et al. 2000; Bennett et al. 2003) expose that the present universe is accelerating and the value of DP lies to some place in the range $-1 \leq q < 0$.\\
\begin{figure}[ht]
\centering
\includegraphics[width=12cm,height=6cm,angle=0]{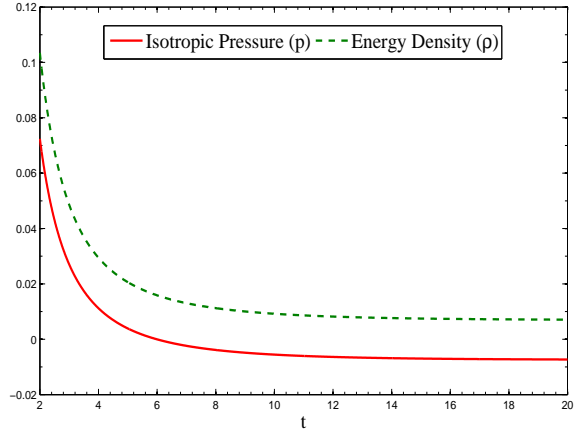}
\caption{The plot of isotropic pressure ($p$) and energy density ($\rho$) vs time ($t$). Here $\alpha = 0.1325$, $n = 2.75$, $m = 0.6$, $c_{2} = 0.00001$.}
\label{fig:figure2}
\end{figure}
\begin{figure}[ht]
\centering
\includegraphics[width=12cm,height=6cm,angle=0]{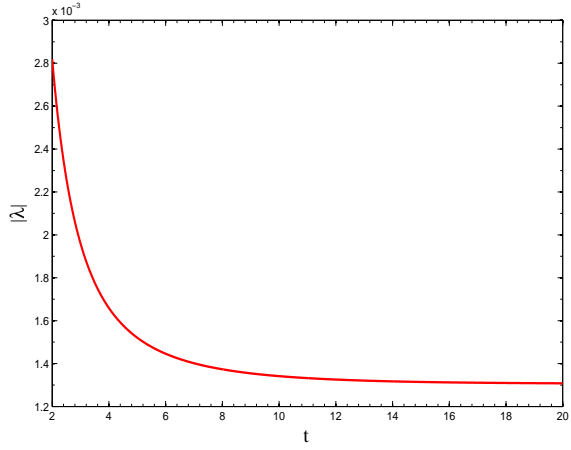}
\caption{The plot of string tension density ($|\lambda|$) vs time ($t$). Here $\alpha = 0.1325$, $n = 2.75$, $m = 0.6$.}
\label{fig:figure3}
\end{figure}
\begin{figure}[ht]
\centering
\includegraphics[width=12cm,height=6cm,angle=0]{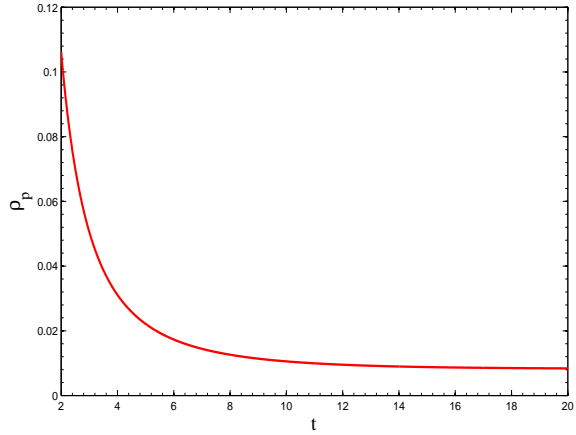}
\caption{The plot of particle energy density ($\rho_{p}$) vs time ($t$). Here $\alpha = 0.1325$,
$n = 2.75$, $m = 0.6$, $c_{2} = 0.00001$.}
\label{fig:figure4}
\end{figure}
\begin{figure}[ht]
\centering
\includegraphics[width=12cm,height=6cm,angle=0]{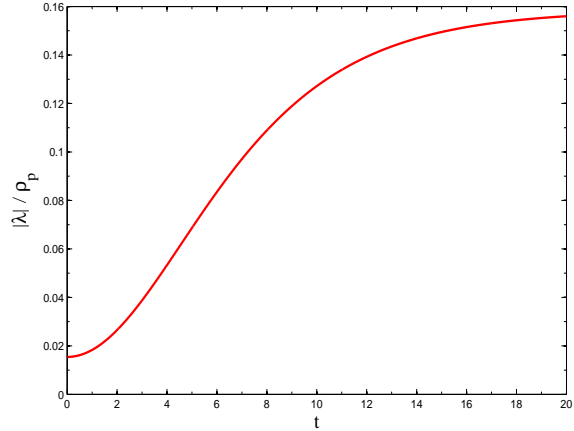}
\caption{The plot of $\left(\frac{|\lambda|}{\rho_{p}}\right)$ vs time ($t$). Here $\alpha = 0.1325$, $n = 2.75$, $m = 0.6$, $c_{2} = 0.00001$.}
\label{fig:figure5}
\end{figure}
Using the values of scale factors from (\ref{eq30})$-$(\ref{eq32}) into the field equations (\ref{eq5})$-$(\ref{eq8}),
we obtain the expressions for string tension density ($\lambda$), energy density ($\rho$) and pressure ($p$) for the
derived model (\ref{eq34}) as
\begin{equation}
\label{eq43} \lambda = \frac{3\alpha^{2}(1-2m)}{2n^{2}(m+1)}\left[(3-n)\coth^{2}(\alpha t) + n\right],
\end{equation}
\begin{equation}
\label{eq44} \rho = \frac{4m+1}{(m+1)^{2}}\left(\frac{3\alpha}{2n}\coth(\alpha t)
\right)^{2} - \left(\frac{c_{2}}{2(\sinh(\alpha t))^{\frac{3}{n}}}\right)^{2},
\end{equation}
\begin{equation}
\label{eq45}
p = -\Biggl[\frac{3\alpha^{2}(8m^{2} + 1)}{4n^{2}(m + 1)^{2}}\coth^{2}
(\alpha t) + \frac{3\alpha^{2}(1 + 2m)}{2n(m + 1)}\left(\frac{1 - n}{n}\coth^{2}
(\alpha t) + 1\right) + \left(\frac{c_{2}}{2(\sinh(\alpha t))^{\frac{3}{n}}}\right)^{2}\Biggr].
\end{equation}\\

\noindent From Eq. (\ref{eq44}), we observe that energy density ($\rho$) of the fluid is a decreasing function of
time and $\rho \geq 0$ when
\begin{equation}
\label{eq46} T^{\frac{3-n}{n}}\sqrt{1+T^{2}} \geq \frac{nc_{2}(m+1)}{3\alpha\sqrt{4m+1}},
\end{equation}
where $T = \sinh(\alpha t)$.\\

\noindent Fig. $2$ depicts the decreasing behavior of isotropic pressure ($p$) and energy density ($\rho$) versus time
($t$). Here we observe that $\rho$ is a positive decreasing function of time and it approaches to zero as $t \rightarrow
\infty$. Moreover, pressure ($p$) varies from early positive values to negative values at late times indicating that the positive values of $p$ resembles for decelerating phase of the Universe while the negative pressure indicates some repulsive
force, which may be responsible for the present accelerating universe which is corroborated by the results from recent SNe Ia observations.\\

\noindent With the help of expressions (\ref{eq43}) and (\ref{eq44}) for the string tension density ($\lambda$) and the
energy density ($\rho$), Eq. (\ref{eq4}) yields the value of particle density ($\rho_{p}$) as
\begin{equation}
\label{eq47}
\rho_{p} = \frac{3\alpha^{2}(8m^{2} + 16m - 1)}{4n^{2}(m + 1)^{2}}\coth^{2}(\alpha t) + \frac{3\alpha^{2}(2m - 1)}{2n(m + 1)}
\left[\left(\frac{1 - n}{n}\right) \coth^{2}(\alpha t) + 1\right] -
\left(\frac{c_{2}}{2(\sinh(\alpha t))^{\frac{3}{n}}}\right)^{2}.
\end{equation}

\noindent From the above expression, we notice that $\rho_{p}$ is a decreasing function of time and it is always
positive when
\begin{equation}
\label{eq48} T^{\frac{3-n}{n}}\left[3(4m^{2}+6m-1)(1+T^{2})-2n(m+1)(2m-1)\right]^{\frac{1}{2}} \geq \frac{c_{2}
n(m+1)}{\sqrt{3}\alpha}.
\end{equation}

Fig. $3$ and Fig. $4$ depicts the decreasing behavior of density parameters ($|\lambda|$ and $\rho_{p}$) versus time ($t$). It is clearly observable from the figures that the particle energy density ($\rho_{p}$) is positive throughout the evolution of the Universe. Moreover, Fig. $5$ gives the plot of $\frac{|\lambda|}{\rho_{p}}$ versus $t$. It is interesting to notice that $\frac{|\lambda|}{\rho_{p}} < 1$ i.e. $\rho_{p} > |\lambda|$ during the cosmic expansion (Kibble 1976; Krori et al. 1990), especially during the early universe indicating that the early universe is dominated by massive strings (Letelier 1983) evolving with deceleration but at later times it got disappear. Since there is no direct evidence of strings at present epoch, therefore, we are interested in the construction of only that models of the Universe which evolves from the era dominated by geometric strings or massive strings, followed by particle dominated era with or without the remnants of strings. Therefore, the derived model describes evolution of the Universe consistent with the present day observations.\\

It is also observed that the above set of solutions satisfy the energy conservation equation (\ref{eq9}) identically. Therefore, the above solutions are exact solutions to the Einstein's field equations (\ref{eq5})$-$(\ref{eq8}).
\section{Physical Validation of the Model}
\begin{figure}[ht]
\centering
\includegraphics[width=12cm,height=6cm,angle=0]{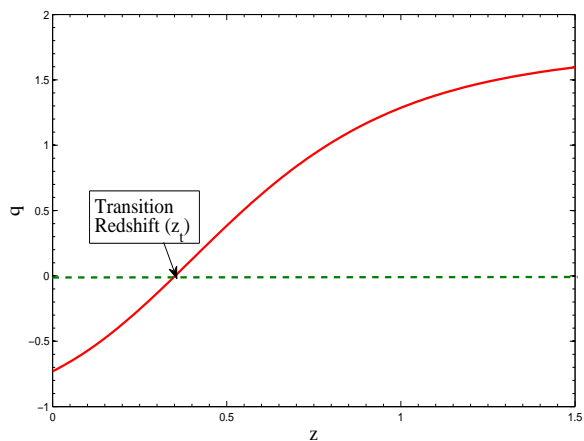}
\caption{The plot of deceleration parameter ($q$) vs redshift ($z$). Here $n = 2.75$ and $q_{0} = -0.73$.}
\label{fig:figure6}
\end{figure}
Here, we firstly define the relationship between average scale factor ($a$) and redshift parameter ($z$) as
\begin{equation}
\label{eq49} z = -1 + \frac{a_0}{a},
\end{equation}
where $a_{0}$ is the present value of the scale factor i.e. at $z=0$.\\

Using the value of scale factor from Eq. (\ref{eq21}) into the above relationship, one can easily obtain
the expressions for cosmic time ($t$), mean Hubble parameter ($H$) and mean deceleration parameter ($q$) in terms of
redshift parameter ($z$) as follows:
\begin{equation}
\label{eq50} t(z) = \frac{1}{\alpha}\sinh^{-1}\sqrt{\frac{n-1-q_{0}}{(q_{0}+1)(z+1)^{2n}}},
\end{equation}
\begin{equation}
\label{eq51} H(z) = \frac{\alpha}{n}\coth\left(\sinh^{-1}\sqrt{\frac{n-1-q_{0}}{(q_{0}+1)(z+1)^{2n}}}\right),
\end{equation}
\begin{equation}
\label{eq52} q(z) = n - 1 - n\left[\tanh\left(\sinh^{-1}\sqrt{\frac{n-1-q_{0}}{(q_{0}+1)(z+1)
^{2n}}}\right)\right]^{2}.
\end{equation}
Such type of relation (\ref{eq52}) provides a two-parameter $(n, q_{0})$ parametrization just like a linear
two-parameter expansion for $q(z) = q_{0} + q_{1}z$ (Riess et al. 2004), where $q_{0}$ is the present value of DP and $q_{1}$ is the deviation in the redshift evaluated at $z = 0$. If we set $q_{0} = -0.73$ (Cunha and Lima 2008), then a positive $z_{t}$ may be obtained only for the positive values of $q_{1}$ since $q_{0}$ is negative and the dynamic transition (from deceleration to acceleration) occurs at $q(z_{t}) = 0$, or equivalently, $z_{t} = -\frac{q_{0}}{q_{1}}$. Another parametrization of considerable interest is $q(z) = q_{0} + q_{1}z(1 + z)^{-1}$ (Xu and Liu 2008) where the parameter $q_{1}$ describes the total correction in the distant past ($z \gg 0$, $q(z) = q_{0} + q_{1}$) and positive $z_{t}$ may be obtained for the positive values of $q_{1} > |q_{0}|$. Also, the dynamic transition occurs at $z_{t} =  -\frac{q_{0}}{q_{0}+q_{1}}$. Moreover, (Xu et al. 2007) have also proposed one such parametrization $q(z) = 1/2 - a/(1+z)^{b}$ where $a$ and $b$ are constants which can be determined from the current observational constraints. As $z \rightarrow \infty$, $q \rightarrow 1/2$ which resembles the value of DP at dark matter dominated epoch. In this case, the present value of DP is $1/2-a$ and $z_{t} = (2a)^{1/b}-1$.\\

For the derived model, the expression for transition redshift ($z_{t}$) is found to be
\begin{equation}
\label{eq53} z_{t} = -1 + \left(\frac{n-q_{0}-1}{(n-1)(q_{0}+1)}\right)^{\frac{1}{2n}}.
\end{equation}

Fig. $6$ shows the variation of $q$ with redshift ($z$) for $q_{0} = -0.73$ and $n = 2.75$. It clearly elaborates
that $z_{t} \cong 0.351817$ which is in fine tuning with (Xu et al. 2007).\\
\subsection{Physical Acceptability of the Solutions}
\begin{figure}[ht]
\centering
\includegraphics[width=12cm,height=6cm,angle=0]{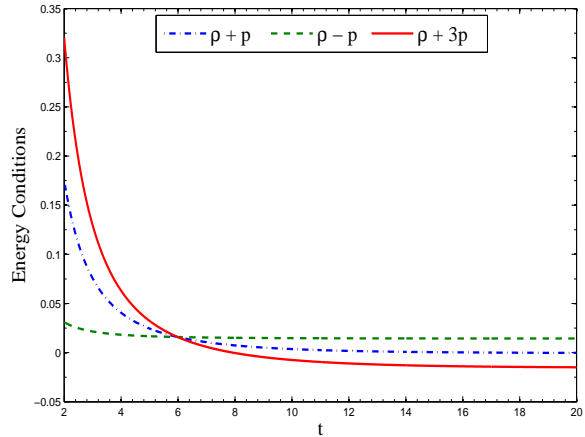}
\caption{The plot of energy conditions vs time $t$. Here $\alpha = 0.1325$,
$n = 2.75$, $m = 0.6$, $c_{2} = 0.00001$.}
\label{fig:figure7}
\end{figure}
Firstly, we check for the fulfillment of energy conditions for the derived model. The weak energy conditions (WEC) and dominant energy conditions (DEC) are given by\\

$(i) \; \rho \geq 0,$ \\

$(ii) \; \rho + p \geq 0,$  \\

$(iii) \; \rho - p \geq 0.$\\

The strong energy condition (SEC) is given by $\rho + 3p \geq 0$.\\

The left hand side of energy conditions have been depicted in Fig. $2$ and Fig. $7$. From both the figures, we observe that
\begin{itemize}
\item The WEC and DEC are satisfied in our model throughout the evolution of universe.

\item The SEC is satisfied during the early stages of the evolution of universe whereas it violates at present epoch. The violation of SEC gives anti-gravitational effect due to which the Universe gets jerk and the transition from the earlier decelerated phase to the present accelerating phase takes place (Caldwell et al. 2006). Hence the presented model clearly describes the late time acceleration of the Universe.
\end{itemize}
\begin{figure}[ht]
\centering
\includegraphics[width=12cm,height=6cm,angle=0]{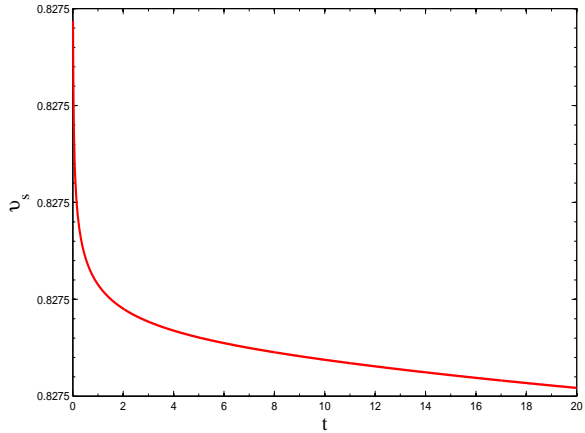}
\caption{The Plot of sound speed ($\upsilon_{s}$) {\it vs.} time $t$. Here $\alpha = 0.1325$,
$n = 2.75$, $m = 0.6$, $c_{2} = 0.00001$.}
\label{fig:figure8}
\end{figure}
Secondly, we check for the physical acceptability: velocity of sound ($\upsilon_{s}$) should be less than the velocity of light ($c$). As we are working in the gravitational units with unit speed of light, i.e., $c=1$, therefore the velocity of sound
($\upsilon_{s}$) must satisfy the condition $0 \leq \upsilon_{s} (=\frac{dp}{d\rho}) \leq 1$.\\

From Eqs. (\ref{eq44}) and (\ref{eq45}), we obtained the sound speed as
\begin{equation}
\label{eq55} \upsilon_{s} = \frac{dp}{d\rho} = \frac{\alpha^{2}\left[2n(m+1)(2m+1)-3(4m^{2}+2m+1)\right](\sinh(\alpha t))^{\frac{6}{n}-2}+nc_{2}^{2}(m+1)^{2}}{3\alpha^{2}(4m+1)(\sinh(\alpha t))^{\frac{6}{n}-2}-nc_{2}^{2}(m+1)^{2}} .
\end{equation}
Fig. $8$ depicts the plot of sound speed ($\upsilon_{s}$) with time ($t$). It is clear form the figure that the speed of
sound is almost constant throughout the evolution of universe and $\upsilon_{s} < 1$.\\

On the basis of above discussions and analysis, we conclude that the derived solutions are physically acceptable. We also emphasize here that the behavior of physical parameters depicted in the figures are due to a selected choice of model parameters while it may change for some different sets of parameters.
\section{Concluding Remarks}
In this paper, exact solutions to Einstein's field equations for spatially homogeneous and anisotropic Bianchi type I model with string fluid as a source of matter and variable DP have been established. The choice of variable DP yields average scale factor $a(t) = (\sinh(\alpha t))^{\frac{1}{n}}$ which represents two types of models: the accelerating models ($0 < n \leq 1$) and the models with transition ($n > 1$) from the early decelerated phase to present accelerating phase (see, Fig. $1$) which is in good agreement with the recent supernovae observations. For the presented model (for $n = 2.75$), the Universe undergoes a phase transition from positive pressure (deceleration) to negative pressure (acceleration). We conclude that this negative pressure plays the role of repulsive force which may be responsible for the current accelerating phase of the Universe. It is also observed that the early universe is dominated by massive strings evolving with deceleration but eventually disappear from the Universe at late time i.e. at present epoch. Moreover, the derived solutions are physically acceptable in concordance with the fulfilment of WEC and DEC and $\upsilon_{s} < 1$ while the violation of SEC is reproducible with current astrophysical observations. Finally, we conclude that the model represents shearing, non-rotating and expanding universe, which starts with a big bang and exhibits point type singularity at $t = 0$. Thus, the solutions demonstrated in this paper may be useful for better understanding of the characteristic of anisotropic Bianchi type I model in the evolution of the universe.

\section*{Acknowledgements}
The authors (AP \& CC) would like to thank the Inter-University Centre for Astronomy and Astrophysics (IUCAA), Pune,
India for providing facility and support during the visit where this work was done. The financial support as J.R.F.
(Sanction No. 09/797(0007)/2010-EMR-I) in part by CSIR, New Delhi, India is gratefully acknowledged by C. Chawla.


\end{document}